# Map Matching based on Conditional Random Fields and Route Preference Mining for Uncertain Trajectories

Xu Ming[①]   Du Yi-man[②]   Wu Jian-ping[②]   Yang Zhou[②]

[①](*School of Computer Science, Beijing University of Posts and Telecommunications, Beijing, 100876, china*)

[②](*School of Civil Engineering, Tsinghua University, Beijing, 100084, china*)

**Abstract**: In order to improve offline map matching accuracy of low-sampling-rate GPS, a map matching algorithm based on conditional random fields (CRF) and route preference mining is proposed. In this algorithm, road offset distance and the temporal-spatial relationship between the sampling points are used as features of GPS trajectory in CRF model, which can utilize the advantages of integrating the context information into features flexibly. When the sampling rate is too low, it is difficult to guarantee the effectiveness using temporal-spatial context modeled in CRF, and route preference of a driver is used as replenishment to be superposed on the temporal-spatial transition features. The experimental results show that this method can improve the accuracy of the matching, especially in the case of low sampling rate.

**Key words**: Map Matching; Conditional Random Fields; Route Preference; Inverted Index

## 1 Introduction

In recent years, the prevalence of GPS enabled devices has led to the production of incredible amounts of vehicle trajectory data, which records human mobility and implies dynamic characteristics of the city. The knowledge discovered in such trajectory data can boost the quality of a variety of novel location-based services and applications, such as pickup location recommendation[1], ridesharing services[2][3], region functions analysis[4], urban planning analysis[5], abnormal events detection[6][7][8], real-time traffic flow prediction[9] and traffic guidance[10][11], etc. In general, a vehicle, equipped with a GPS sensor, reports its time-stamp locations to a data center periodically. However, the readings of GPS sensor frequently deviate from the actual positions due to measurement errors of the device and the effect of communication environment. Therefore, map matching, the process of lining up all observed points of a trajectory on correct road segment, is key pre-process step in these trajectory mining applications and given considerable attention by researchers. Besides, different from requiring a short response time in real-time route navigation, map matching in trajectory mining applications emphasizes on high accuracy, so it is also called offline map matching.

With respect to offline map matching, a great challenge is that we have to be confronted with the uncertainty caused by low-sampling-rate, apart from sampling errors and positioning errors. In fact, for energy consumption considerations, GPS sampling interval of most vehicles are more than 30 seconds, and even a considerable number exceed 2 or 3 minutes. If a vehicle travels on the road network in a city with a free-flow speed and its GPS sampling interval is 3 minutes, the distance between two neighborhood sampling points can reach to approximate 3km. Some significant neighborhood context information used in improve the accuracy is lost in such low frequency trajectories. This uncertainty increases difficulty of identifying the real road, on which the vehicle is travelling. Consider an example in Figure 1, a trajectory of a vehicle with 20 seconds sampling interval is presented, comparing with its sparse variety with 3 minutes sampling intervals (removing some sampling points of the original trajectory). Obviously, the path of the high-sampling-rate trajectory can be identified clearly; while the path of the sparse trajectory is difficult to find out.

In this paper, we explore a way of reducing effect of uncertainty on map matching to improve the effectiveness. To achieve this, we need to make full use of available information, such as characteristics of trajectory, topology of road network and historical route mode of drivers. We treat offline map matching of GPS sampling points as a sequence labeling problem in machine learning, and propose a conditional random fields (CRF) [12] map-matching algorithm, which employs the spatial and temporal correlation between neighborhood sampling points. There also exist other probabilistic approaches suitable for sequence labeling problem, such as Hidden Markov Model (HMM) [21] and Maximum Entropy Markov Model (MEMM) [13]. However, due to the structure, HMM model needs assuming that all output observations are conditionally independent and in the process of state transition, current state only depends on the previous state. This may loses some significant context, and results in a labeled bias problem that the model inclines to match sampling points to long-distance road segment linked less intersections. Although MEMM avoids the conditional independence assumptions on observations by improving structure of probabilistic graphical model, it also suffers from labeled bias problem. Compared with HMM and MEMM, CRF overcomes the drawbacks in them and has obvious advantage on flexible integration of a variety of features and context information. In addition, we find that most drivers usually choose familiar paths to travel. Inspired by this, we attempt to employ route preferences of drivers to better the quality of the algorithm. Specially, if GPS sampling frequency of given trajectory is too low to guarantee effectiveness of CRF features, we exact the route preferences from matched paths of historical trajectories, and weighted superpose them on the outcome of CRF. In fact, with respective to a trajectory with high uncertainty, our algorithm is inclined to match the GPS observations to the roads on the familiar paths of the driver. Finally, experiments verify the valid of this way.

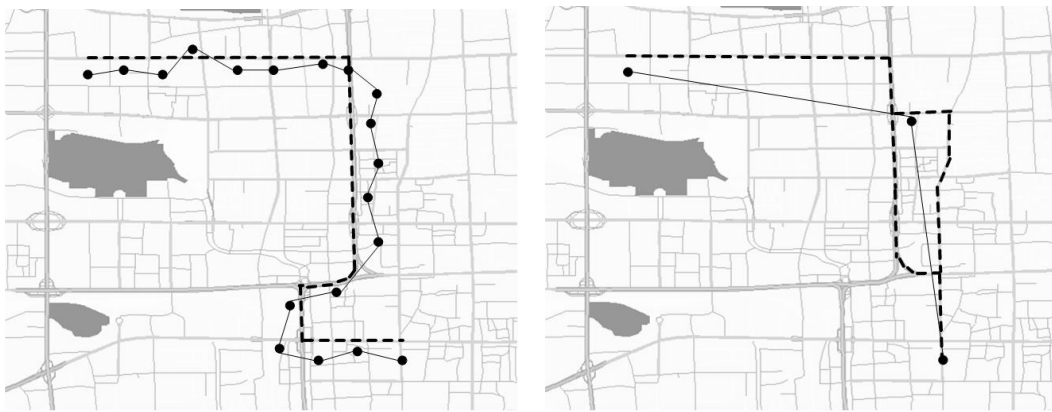

Fig.1 A comparison of map matching high-sampling-rate trajectory (left) and low-sampling-rate trajectory (right)

In summary, this paper makes the following contributions:

1) We propose a CRF-based algorithm of map matching, and it combines both the spatial and temporal features effectively. Experimental results show that the temporal features can boost discriminating power of existing CRF model that solely employs spatial feature, especially in some areas with similar spatial features, such as interaction-dense areas.

2) We design a framework of mining route preferences of drivers from historical matched trajectories to improve effectiveness of CRF-based algorithm, and it can weaken the effect of the uncertainty causing by low-frequency-sampling.

3) We perform extensive experiments on real trajectory dataset collected from the physical world and labeled manually. Our algorithm is evaluated on matching accuracy. The results show that our algorithm outperforms previous methods significantly for low-sampling-rate trajectories.

The remainder of the paper is organized as follows. A review of some existing map matching methods is given in Section 2. Section 3 presents our map matching framework and problem definition. Then the algorithm is proposed with detailed discuss and analysis in Section 4. Section 5 presents the experimental evaluation. Finally, we conclude this paper in Section 6.

## 2 Related Work

Map matching problem attracts lots of researchers and a large number of approaches are proposed. According to application scenario, these methods can be divided into two categories: online and offline.

Online algorithms mainly develop a greedy strategy to search for the local optimal matching from an already solution. Greenfeld [14] proposes an incremental algorithm that only considers the geometric information to evaluate each candidate edges. This information contains the distance similarity and the orientation similarity. Chawathe [15] proposes a segment-based method, in which the confidence score is defined and assigned to different sampling points. When a new trajectory comes, the high confidence edges are matched first, and then low confidence edges are matched according to already matched edges. Wenk et al. [16] propose an "adaptive clipping" approach that obtains the shortest path on local free space graph. This kind of methods can figure out the matching segments with short response time due to using only a small part of the trajectory, so they are widely used for on-line applications, such as navigation system. However, the accuracy of these algorithms falls down sharply when sampling frequency decreases.

Offline algorithms handle the entire trajectory after completing a trip. Most of studies detect the closest candidate roads from the current trajectory by means of Fréchet distance or its extended metrics, its underlying meaning is that the continuity of curve is taken into account to search for corresponding paths. In the algorithm proposed by Alt et al. [17], the critical values are worked out in a parametric search process, and then Fréchet distance is measured by finding a monotone path in the free space from the lower left corner to the upper right corner. In order to reduce the effect of the anomalous sampling points in this work, Brakatsoulas et al.[18] propose an extended algorithm using average Fréchet distance, in addition, weak Fréchet distance is used in their work to reduce the time cost to $O(mn \log mn)$. Yin et al [19] model road network using the weight graph, and their proposed algorithm is based on edit distance, which is similar to Average Fréchet distance. However, these deterministic algorithms are also susceptible to noise, and perform worse in low sampling rates.

To deal with the noise, low sampling rate and other issues effectively, methods based on probabilistic are widely used. Lou et al. [20] propose an ST-Matching algorithm that combines temporal and spatial context. At first, the candidate roads of each sampling point in the given trajectory are determined according to Euclidean distance between current point and each candidate road, and then use spatial and temporal analysis to calculate observation probability and transition probability. After accumulating probability scores, the path, which has the maximum joint probability, would be considered as the matched path. However, this method does not take into account the weights of different factors and interaction between non-neighbor points, so the accuracy falls rapidly when the path is too long or in the area existing multiple lanes. Newson and Krumm [21] propose a map-matching algorithm for low-sampling-rate trajectories based on hidden Markov model (HMM). In their work, observation probability matrix and transition probability matrix are inferred by learning on the training dataset, and then Viterbi algorithm [12] is used to get the result. But HMM has the too strict independence assumption, ignoring impact between points with a long distance and non-orthogonality of features, so its accuracy is slightly lower than ST-Matching. Liao et al. [22] also give a CRF-based algorithm, but since it only employ the spatial context, the sampling points are inclined to match to roads on the shortest path, and it is not suit for low-sampling-rate trajectories.

## 3 System Overview

### 3.1 Problem definition

In this section, we give definitions of some terms used in this paper.

Definition 1: A **trajectory** $\theta$ is a sequence of time-ordered GPS points generated by a vehicle in a trip. It can be represented by a two-tuples $<v, o>$, where $v$ denotes the identifier of a vehicle, $o$ denotes the sequence of GPS points, $o = o^{(1)} \rightarrow o^{(2)} \rightarrow ... \rightarrow o^{(T_h)}$, $T_h$ is the total number of sampling points in the $h^{th}$ GPS trajectory. $o^{(1)}$ and $o^{(T_h)}$ denote origin and destination respectively. Each GPS point $o^{(t)}$ can be represented by three-tuples $<x, y, t>$, where $x$ denotes latitude, $y$ denotes longitude and $t$ denotes timestamp. Let $\Theta$ denotes the collection of $N$ trajectories.

Definition 2: a **trip** $tr$ is a pair of origin-destination (OD), which is represented by a two-tuples $<r_o, r_d>$, where $r_o$ denotes origin road and $r_d$ denotes destination road. If origin and destination road of two trips $tr_A, tr_B$ are the same respectively, $tr_A = tr_B$ can be considered tenable.

Definition 3: a **path** $\gamma$ is a sequence of road segments traversed by a vehicle in one trip. $\gamma = r^{(1)} \rightarrow r^{(2)} \rightarrow ... \rightarrow r^{(T)}$, where $r^{(1)}$ denotes origin road, $r^{(T)}$ denotes destination road. $r^{(t)} \in \{r_w\}_{w=1}^{W}$, here $r_w$ is the identifier of a road segment, $W$ is total number of the road segment in the road network. If any two neighborhood road segments of a path are different and topologically connected, such path is called complete path.

Map matching is equivalent to sequence labeled problem in machine learning. Given an observable GPS trajectory $\theta$, sampling points sequence $o$ can be regarded as an observation sequence to be labeled; road set $\{r_w\}_{w=1}^{W}$ can be regarded as the label set. The objective is to find a path $\gamma^*$ optimal matching for trajectory $\theta$, $\gamma^*$ is essentially maximum a posteriori probability path, i.e. $\gamma^* = \arg\max_{\gamma} p(\gamma | o)$.

3.2 System framework

The framework of our proposed algorithm is presented in Figure 2. It consists of training phase and prediction phase. In the training phase, the parameters of CRF model are inferred by learning on labeled trajectories. In prediction phase, at first, observation features and transition features are extracted from the given trajectory and computed to determine CRF model, and then the value of average sampling interval $f$ is checked, if $f$ is not lower than a given threshold, the matched path is calculated using CRF model directly. In detail, conditional probability of road segment given any sampling point equals the sum of all the features, and then the matched path with maximum joint conditional probability is worked out. Otherwise, the sampling rate is considered too low to capture the temporal and spatial correlation, in order to improve matching accuracy, route preference information is extracted from historical trajectories of the corresponding vehicle, and then it is weighted superposed on transition features of CRF to generate new transition features.

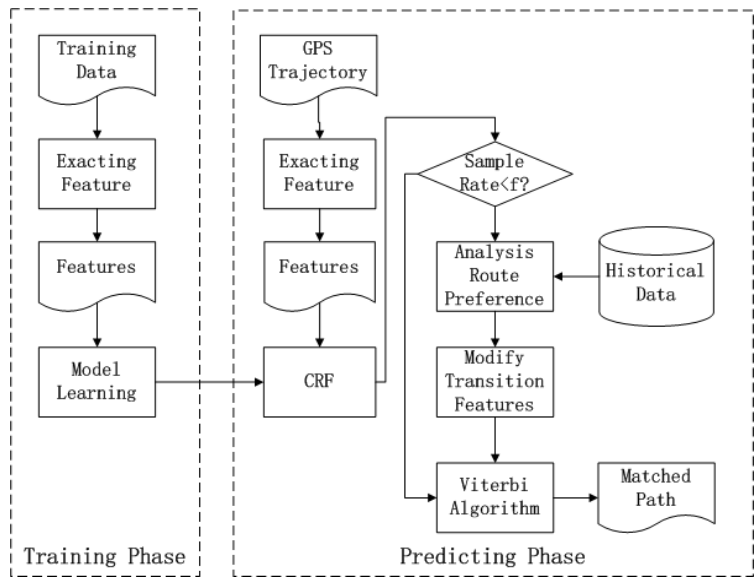

Fig.2 The framework of proposed map matching algorithm

4 Proposed Approach

4.1 CRF Based approach

As a special case of undirected graphical models, conditional random field is developed on the basis of the maximum entropy model, and is widely used in sequence labeled problem. It avoids the labeled bias problem in the maximum entropy model.

Definition 4: Let $X$ and $Y$ be two sets of random variables, $p(Y|X)$ is conditional probability of $Y$ given $X$. If the set $Y$ of random variable constitutes an undirected graph $G=(V,E)$ satisfying the Markov property, that is to say, $p(Y_v|X,Y_w, w \neq v) = p(Y_v|X,Y_w, w \sim v)$ is founded on any node $v$ in graph, where $w \sim v$ denotes all the nodes connecting node $v$ in graph $G$, $w \neq v$ denotes all the nodes except node $v$. The probability distribution $p(Y|X)$ is called condition random fields.

Sequence labeled problem can be model using a linear CRF. If we denote the values assigned to observation sequence $X$ by a vector $x$, the conditional probability of assigning vector $y$ to labeled sequence $Y$ has the following form

$$p(y|x) = \frac{1}{Z(x)} \exp\left( \sum_{i+1 \in V} \left( \sum_j \varphi_j(y_i, x, \mu_j) + \sum_k \delta_k(y_i, y_{i+1}, x, \lambda) \right) \right) \quad (1)$$

Here, $Z(x) = \sum_y \exp\left( \sum_{i+1 \in V} \left( \sum_j \varphi_j(y_i, x, \mu_j) + \sum_k \delta_k(y_i, y_{i+1}, x, \lambda) \right) \right)$ is the normalized factor, which is used to convert $p(y|x)$ to the normalized probability. $\varphi(\bullet)$ is the potential function defined corresponding the nodes in the graph model, and represents the possible of a certain label appearing without considering the effect of other labels, it is also called generative feature. $\delta(\bullet)$ is the potential function corresponding the edges linking nodes. It indicates the dependencies of labels and is also called transition feature. $\mu$ and $\lambda$, parameters of CRF, control the weights of potential functions. The local information of graph is modeled using potential functions, and such context information can be propagated through the edges linking the nodes; therefore CRF can integrate a wide range of context information.

With respect to map matching, the road segment of current sampling point may be depend on the road segments, which the vehicle is on a few sampling intervals earlier. In theory, the high order CRF, which integrates the long distance context dependencies, should be used to model to achieve the high accuracy. However, the parameters inference of high order CRF needs computing in a more complicated process. In this paper, we use a simple first order linear CRF model, in which the dependency between labels of neighborhood sampling points is solely considered. Our experiments verify the valid of modeling in this way. In our model, the generative feature is denoted by $\varphi(r^{(t)}, o^{(t)})$, which represents the probability that a GPS observation $o^{(t)}$ is observed when the vehicle travels on road segment $r^{(t)}$ at sampling time $t$. The transition feature $\delta(r^{(t)}, r^{(t+1)})$ represents the probability that the vehicle is on the road segment $r^{(t)}$ at sampling time $t$, and traverse on the road segment $r^{(t+1)}$ at next sampling time $t+1$. Therefore the GPS sequence $o$ is given, the posterior probability of the matching path $\gamma$ can be calculated with

$$p(\gamma|o) \propto \exp\left( \sum_{t+1 \leq T} \mu \varphi(r^{(t)}, o^{(t)}) + \lambda \delta(r^{(t)}, r^{(t+1)}) \right) \quad (2)$$

4.2 feature selection

In this section, we give the concrete expression of each feature function. According to mentioned earlier, generative feature is only determined by the distance between the GPS point and its candidate road segment. Obviously, the road segment, which is closer from the observed point, has greater possibility of generating the observation. We model the error of GPS point using a Gaussian distribution $N(0, \sigma^2)$, the formula of generative feature is given by

$$\varphi(r_m^{(t)}, o^{(t)}) = \frac{1}{\sqrt{2\pi}\sigma} \exp\left(-\frac{d_p^2(r_m^{(t)}, o^{(t)})}{2\sigma^2}\right) \quad (3)$$

Here, $r_m^{(t)}$ denotes $m^{th}$ candidate road segment of GPS observation $o^{(t)}$ at sampling time $t$, $d_p(r_m^{(t)}, o^{(t)})$ denotes the projection distance from $o^{(t)}$ to $r_m^{(t)}$. Further, the standard deviation $\sigma$ and the constant coefficient can be incorporated into a parameter $\mu$. Then we get the simple generative feature given as follow:

$$\varphi(r_m^{(t)}, o^{(t)}) = \mu \exp\left(-d_p^2(r_m^{(t)}, o^{(t)})\right) \quad (4)$$

In fact, although $\varphi(r_m^{(t)}, o^{(t)})$ does not follow Gaussian distribution strictly, the experimental results show that the effect of the model is ideal, since the goal of feature extraction is to enable the model to have a greater discriminating power in the feature space. The transition features model the possibility of jumping between the candidate road segments of neighborhood points. In general, drivers would give priority to choose the shortest route, and the candidate road segments of neighborhood GPS points should be adjacent or chose to one another in spatial topology. Therefore, the spatial transition feature $\delta_1$ is defined as

$$\delta_1(r_m^{(t)}, r_n^{(t+1)}) = \frac{d^2(o^{(t)}, o^{(t+1)})}{d_r^2(e_m^{(t)}, e_n^{(t+1)})} \quad (5)$$

Where $e_m^{(t)}$ is the projection point of GPS observation $o^{(t)}$ on the $m^{th}$ candidate road segment. $d(\bullet)$ and $d_r(\bullet)$ are functions for calculating the Euclidean distance and path distance between two points respectively. Obviously, $\delta_1(\bullet) \in (0,1)$, and the smaller value of $\delta_1(\bullet)$, the more detour the path. $\delta_1(\bullet)$ reflects the spatial dependency of neighborhood observations. However, in some cases, the spatial transition feature cannot be reliable sufficiently to determine how to distinguish the actual path from the other candidates, e.g., in some dense parts of road network, the spatial features of candidate road segments may be similar. In order to guarantee the performance of algorithm, other high discernable feature should be integrated into our model. Consider that the speed constraint of a road segment may be different from others, like expressway and bypass; even with respect to the same road, the average speed is quite different in different time. It is not possible to match the GPS observation to the road segments, on which the vehicle exceeds the speed limit. Thus, we introduce the temporal feature. Assume that there are $k$ road segments between $e_m^{(t)}$ and $e_n^{(t+1)}$, the historical average time $\Delta t_e$ of traversing from $e_m^{(t)}$ to $e_n^{(t+1)}$ can be calculated by $\Delta t_e = \sum_k \frac{r_k.l}{r_k.v}$, where $r_k.l$ denotes the distance of traversing on road segment $k$, $r_k.v$ denotes historical average speed of road segment $k$ at the same time slot. The temporal transition feature $\delta_2$ is given as

$$\delta_2(r_t^m, r_{t+1}^n) = \left(\frac{\min(\Delta t, \Delta t_e)}{\max(\Delta t, \Delta t_e)}\right)^2 \quad (6)$$

Like with $\delta_1(\bullet)$, $\delta_2(\bullet) \in (0,1)$. Due to taking temporal feature as consideration, model would give a higher probability of matching GPS point to the candidate road, whose average speed is closer to the speed of the trajectory.

In summary, transition feature $\delta$ is weighted sum of spatial feature $\delta_1$ and temporal feature $\delta_2$. It is given by

$$\delta(r_m^{(t)}, r_n^{(t+1)}) = \lambda_1 \cdot \delta_1(r_m^{(t)}, r_n^{(t+1)}) + \lambda_2 \cdot \delta_2(r_m^{(t)}, r_n^{(t+1)}) \quad (7)$$

The weighted coefficients $\lambda_1$ and $\lambda_2$ are also the model parameters, which are determined in training phase.

4.3 Model Inference

The goal of model inference is to infer parameters $\omega = \{\mu, \lambda_1, \lambda_2\}$, which is independent on time $t$. The training set containing labeled trajectories is needed in this phase. Given the definition of conditional probability $p(\gamma | o)$, optimization goal is to maximize likelihood of the training data, the logarithm likelihood function is given by $\ell(\omega) = \sum_l \log p(\gamma^{(l)} | o^{(l)}; \omega)$. A standard parameter learning process is to calculate the gradient of the objective function $\ell(\omega)$, then utilize this gradient to search for optimal solution. There are many algorithms to complete this task, and we use L-BFGS [23] in this paper.

In the actual process of parameter learning, in order to prevent the candidate road space from becoming too large, require that maximum number of candidate road segment corresponding to each GPS observation is six. In other words, given the location of GPS point, top 6 nearest road segments are selected as candidates. Maximum likelihood estimates of $\omega$ can be obtained after parameters learning, it is denoted by $\hat{\omega}$. Map matching is the process of seeking the maximum a posteriori estimation of $p(\gamma | o, \omega)$. In this paper, we use Viterbi algorithm [12], which can calculate global optimal solution with lower time complexity.

4.4 Map Matching Based on CRF and Route Preference Mining (RPM)

CRF model above can already match a GPS trajectory to the corresponding path. However, dealing with lower-sampling-rate trajectory, accuracy of the algorithm is still not satisfactory. The main reason is that the correlation between neighborhood observations decreases with the sampling intervals increasing, especially in the zone of dense interactions, the spatial feature and temporal feature of a candidate road segment are similar to others. Consider the example in Figure 3, it is difficult to distinguish the path of trajectory $o^{(t-1)} \to o^{(t)} \to o^{(t+1)}$ between $r_1 \to r_2 \to r_5 \to r_6$ and $r_1 \to r_3 \to r_4 \to r_6$. To tackle this issue, route preference information is introduced to supply the lack of discrimination of existing features. In previous research, Froehlich et al. [24] find that 60 percent of paths of a driver are repeated, and most drivers select routes following their personal preferences, moreover, a confluence of paths selected by similar preferences drivers may generate a popular route. We attempt to discover and quantify personal route preference, and then this information them to improve the performance of CRF-based algorithm.

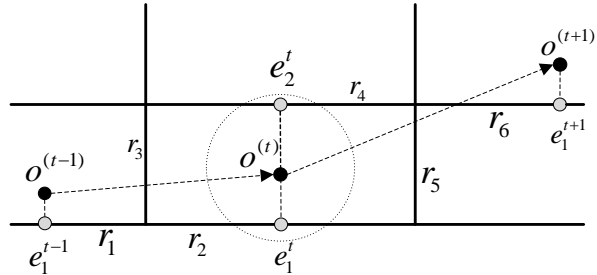

Fig.3 Illustration of difficulty of determining which path should be matched to the low sampling frequency trajectory

Firstly, for each observation in a trajectory of vehicle $v$, a 6-candidate road segment collection is obtained by calculating top-6 shortest distance, and then the possibility of jumping between two candidate roads of neighborhood observations is obtained using statistical method in historical matched trajectories. This possibility is regarded as the route preference which is denoted by $h_v(r_m^{(t)}, r_n^{(t+1)})$, and which can be calculated by

$$h_v(r_m^{(t)}, r_n^{(t+1)}) = f_v(r_m) p_v(r_n^{(t+1)} | r_m^{(t)}) \tag{8}$$

where $p_v(r_n^{(t+1)} | r_m^{(t)})$ represents the conditional probability of traversing on $n^{th}$ candidate road segment at time $t+1$, given condition of a vehicle travesing on the $m^{th}$ candidate road segment at time $t$, and this can be calculated by

$$p_v(r_n^{(t+1)} | r_m^{(t)}) = \frac{p_v(r_n^{(t+1)}, r_m^{(t)})}{p_v(r_m^{(t)})} = \frac{\text{Count}_v(r_m \to r_n) + 1}{\sum_{j=1}^{I_{t+1}} \text{Count}_v(r_m \to r_j) + 1} \tag{9}$$

Here, $\text{Count}_v(r_m \to r_n)$ denotes the number of path, which contains the road segments $r_m$ and $r_m$ following by order $r_m \to r_n$. Calculating $p_v(r_n^{(t+1)} | r_m^{(t)})$ is entirely dependent on statistical methods, so if the samples are scarce, it cannot reflect actual probability of moving between road segments. To be more accurate, $f_v(r_m)$, which represents driving experience of driver $v$ on road $r_m$, is introduced. The growth of driving experience $f_v(r_m)$ can be modeled using a sigmoid curve as follow

$$f_v(r_m) = \frac{1}{1+e^{-(ax_m^v+b)}} \tag{10}$$

Where $x_m^v$ is number of times of vehicle $v$ traversing on $r_m$. $ax_m^v + b$ is the linear transformation mapping $x_m^v$ from $[0,+\infty]$ to $[-5,+5]$, $a$ and $b$ are coefficients. Obviously, the more times the vehicle $v$ traverses on road $r_m$, the closer to 1 $f_v(r_m)$ is, which indicates that if driver $v$ is more familiar to road $r_m$, the route preference makes greater contribution to prediction results. After the route preference is figured out, it is used to superposed on the transition features of CRF, and we can get the new transition probability given by

$$s'(r_t^m, r_{t+1}^n) = \alpha \cdot h_i(r_t^m, r_{t+1}^n) + (1-\alpha) \cdot s(r_t^m, r_{t+1}^n) \tag{11}$$

Here $\alpha$ is weighted factor, which is used to control the portion between features in CRF and route preference, and which is determined by experiment.

When calculating $h_v(r_m^{(t)}, r_n^{(t+1)})$, queries for number of a certain road or a certain path passed by a vehicle are required frequently. To guarantee a short response time, we design an inverted-index table (IDT), which is implemented in three layers of nested hash structure. Its structure is showed in Figure 4. The key of first layer hash represents each vehicle ID, and its value is a pointer, which point to second layer hash. The key of second layer hash is road segment ID visited by corresponding vehicle, and its value is also a pointer pointing to the third layer hash table. The key of the third layer is ID of path, which includes the road segments traversed by corresponding vehicle, and the value of third layer is the road segment order in the corresponding path. It is used to determine the direction of the path.

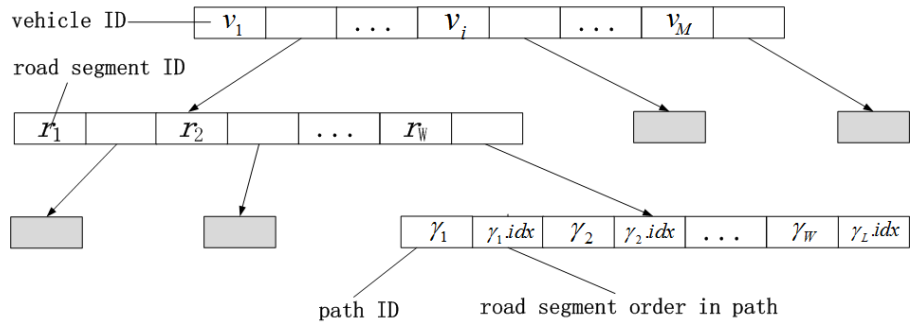

Fig.4 The structure of inverted-index table

In general, the drivers adopt different route strategies in different times. For example, a detour is taken to guarantee the minimum travel time in morning peak hours, and the shortest path or accustomed route is selected in normal times. Therefore, in order to calculate route preference accurately, we partition a day into 3 time slots: morning peak ranges from 7:30 to 9:30, evening peak ranges from 17:30 to 19:30, and normal period is rest time of peak hours. Each IDT is built respectively in each time slot. When launching a query, the request is handled in corresponding IDT according to the sampling time. Table 1 shows our proposed algorithm combining CRF and route preference.

## 5 Experiments

### 5.1 Experimental Setting and Dataset Description

For our experiments, we construct a LAN consisting of three computers, which are used to provide GIS service, database and execute the algorithm respectively. We implement our algorithm using C#, GIS server is set up with ArcGIS 10. The digital map mainly includes road network layers with "shape" format. The attributes of road network include ID, name, level, and length. The traffic flow information of network is derived from statistics on historical trajectory data. All information, such as original GPS, traffic flow, the path of each trip and the IDT, is stored in SQL Server.

Table 1 Map matching algorithm based on CRF and route preference mining

| | |
|---|---|
| Input: a trajectory $o_{1:T}$ of vehicle v, CRF model, IDT of vehicles | |
| Output: path (road segment sequence) | |
| 1 | Check the average sampling interval of trajectory, if lower than threshold, **goto** 2; **otherwise goto** 8; |
| 2 | **for** t:=0 to T-1 **do** |
| 3 |    **for** each candidate road i in candidate road set of GPS observation of time t **do** |
| 4 |       **for** each candidate road j in candidate road set of GPS observation of time t+1 **do** |
| 5 |          Count the number $C_{i->j}$ of paths passing through i to j in order and the number $C_i$ of paths passing through i |
| 6 |          Calculate the route preference of driver v using formula (8) |
| 7 |          Calculate new transmission features using formula (11) |
| 8 | Calculate the maximum posterior probability using Viterbi algorithm. |

We use real trajectory data set generated by 720 Beijing taxis in a period of 6 months (from March 2012 to August 2012). The sampling interval is approximate 10 seconds. The entire data set is labeled manually. To validate the regularity of the personal route, we define the repeated path $\gamma_c$ of path $\gamma_h$, which can be computed as

$$repeat(\gamma_c \rightarrow \gamma_h) = \frac{\text{Card}(\gamma_c \cap \gamma_h)}{\text{Card}(\gamma_c)} \quad (12)$$

When $repeat(\gamma_c \rightarrow \gamma_h)$ is greater than the threshold $\zeta$, the path $\gamma_c$ can be considered as repeated path of $\gamma_h$, here $\zeta$ is set to 0.8. Whereas, note that, $\gamma_h$ is not necessarily the repeated path of $\gamma_c$. For example, $\gamma_h$ is a path containing ten road segments, and $\gamma_c$ is sub-path of $\gamma_h$, which contains four road segments. The statistic of our dataset is showed in Figure 5. We find that the ratio of repeated path increases with the growth of amount of data, and the taxi routes show the obvious repetitiveness. When amount of trajectories is accumulated to two months, average 32.4% of paths are repeated. For 68 of these vehicles, the rate is 50%+. Only 47 vehicles have less than 20% repeated paths, since their daily trips are fewer than others, When data amount increases to five months, the average of repeated paths rate grows to 63.2%, and 54 vehicles have 80%+ repeated rate and the minimum of them reaches 40.4%.

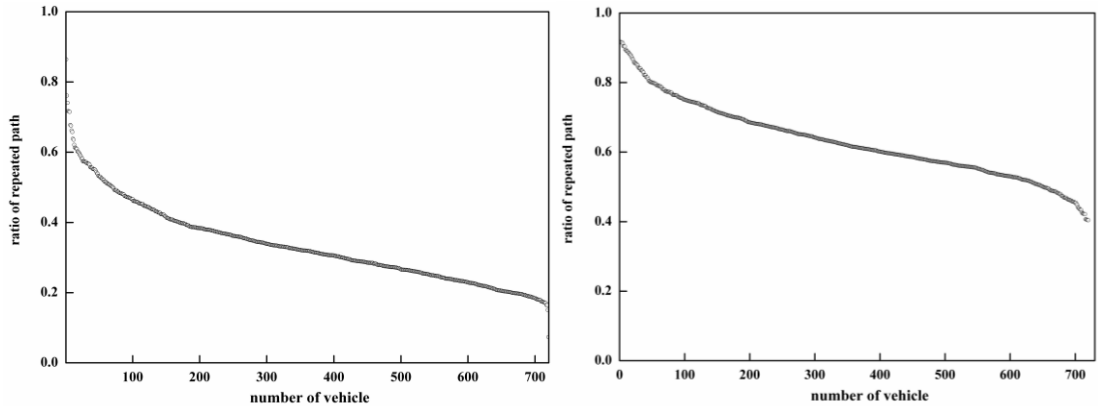

Fig 5 comparison of the repeated paths ratio in two months (left) and five months (right) of trajectories dataset

We also validate that the number of repeated path grows with more days of observation. Figure 6 (left) presents the growth trend of one driver's repeated path ratio over a period of five months. At first, the ratio increases relative slowly, and it is less than 20% when accumulating to 40$^{th}$ day, which indicates that more obscure paths are generated in this period. Then the growth rate increases significantly with the number of days continues to grow, when accumulating to 90 days, average of repeated path ratio is 60%+. Thereafter, the growth rate seems to be slow again, finally the average of this ratio is held at approximate 70%. Further, the regularity of route selection is verified through statistical distribution of paths. Figure 6 (right) presents 387 paths derived from 6820 trips of a driver over 5 months. We find that the major path bears 235 trips, which occupy 3.4% of total number; moreover, 1432 trips, 21% of total, are distributed on the top 10 frequent paths. This inhomogeneity of paths distribution confirms the existence of personal route preference. Therefore, the utilization of path selection knowledge can effectively supply the missing context information caused by the low-sampling-rate, and improve predictive accuracy.

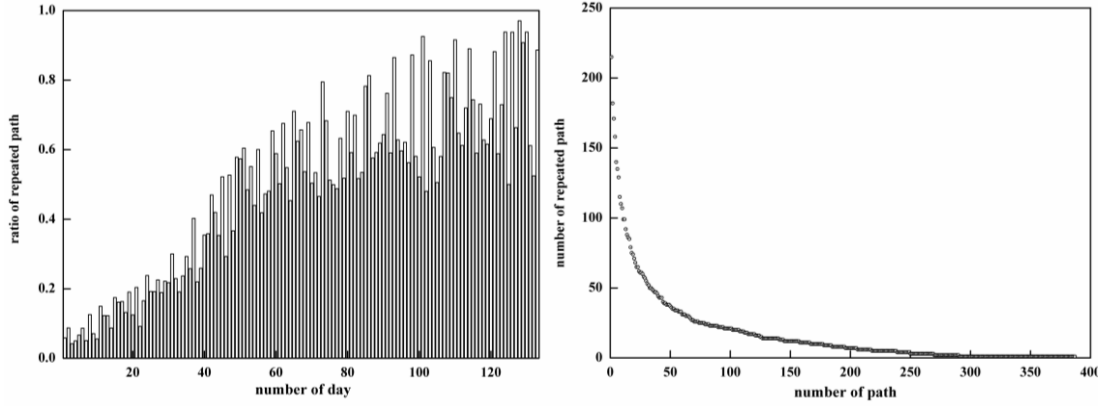

Fig.6 the relation between repeated paths ratio and days, (left) and distribution of number of repeated paths (right) for a taxi

5.2 Experimental Result

The effectiveness of our proposed algorithm is evaluated using two accuracy metrics: Accuracy by road segments ($A_s$) and Accuracy by paths ($A_r$) are defined as

$$A_s = \frac{\#\text{correctly matched road segments}}{\#\text{all road segments of the trajectories}}$$

$$A_r = \frac{\#\text{correctly matched paths}}{\#\text{all paths of the trajectories}}$$

Firstly, we estimate the effects of our CRF model ($CRF^1$) by comparing with incremental, HMM and CRF in [22] ($CRF^2$). We select any five months trajectories as training set, and process the remaining one month trajectories to generate a plurality of groups at different time intervals (by 30s increments) for testing. Figure 7 shows the results of each algorithm.

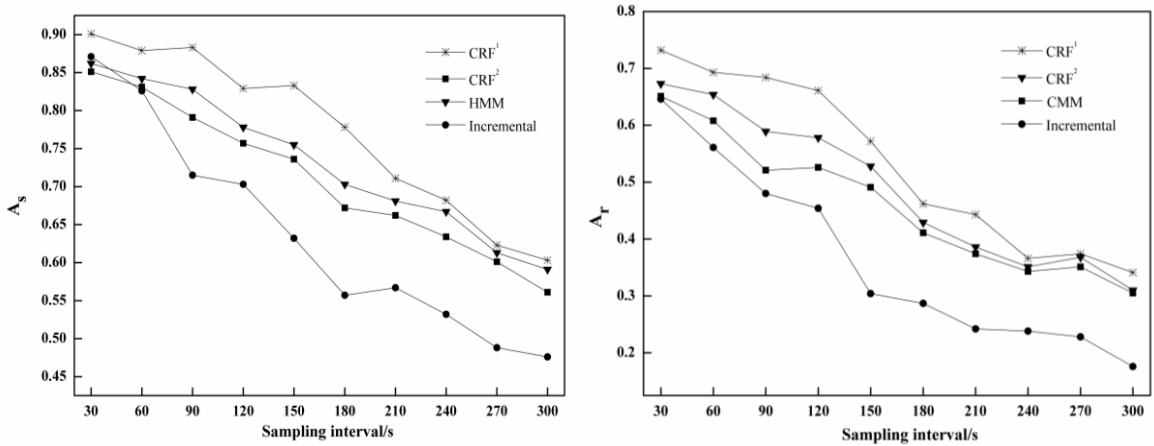

Figure 7 Comparison of HMM, Incremental algorithm, $CRF^1$ and $CRF^2$

As demonstrated in figure 7, under the same conditions, $A_r$ is much lower than $A_s$ due to the strict restriction of $A_r$. When the sampling interval is less than 30s, all algorithms can achieve high accuracy. And with the sampling interval increases, the accuracy of all algorithms shows various degree of downward trend. Because all of these algorithms use the context information of the sampling points, and the effects of context information are weakened gradually with sampling frequency decreasing. Specifically, the accuracy of incremental approach using the local geometric features drops more dramatically. Both HMM and $CRF^2$, which only consider the spatial context, accordingly, exhibit the similar performance. And the accuracy of $CRF^2$ is slightly higher, since $CRF^2$ model obtains the optimized weights of each feature by parameter learning and makes better use of the spatial context. As temporal correlation is considered, when the sampling interval is not too large, $CRF^1$ significantly outperforms $CRF^2$. These correct matching resulted from temporal constraints are concentrated in the trips that mainly happen in the period of morning or evening peak hours. During these peak hours, probably, the shortest path is not the fastest path fastest path due to traffic congestion. In addition, in some regions of dense and diverse roads, many candidate paths of GPS observations have the similar spatial context information, thus the temporal correlation can heighten the divergence among the solutions to some extent. We also can find that, with sampling interval continues to increase, the accuracy curve of $CRF^1$ drops sharply and gets close to $CRF^2$, which indicates that temporal correlation has weaker effect when neighboring observations are far away from each other.

The parameter $\alpha$ in our proposed framework is used to balance the ratio of temporal-spatial constraints and route preference. We estimate $\alpha$ through analyzing accuracy change, which is caused by setting $\alpha$ to different values. We use any five months matched trajectories to calculate route preference, and set up three groups of test sets according to different sampling interval: 180s, 300s and 420s. Figure 8 shows the evaluation results. In test set on 180s interval, with the value of $\alpha$ rising, $A_s$ grows gradually first, and reaches the maximum 0.821 when $\alpha$ is 0.7, then $A_s$ shows a decline trend. The reason is that when the proportion of route preference is too high, its effect can covers the effective temporal-spatial features. Although, some GPS observations are matched to the correct road segments, more observations are mismatched to road segments, on which the corresponding driver traverses frequently. Meanwhile, $A_r$ exhibits a tendency toward stabilization after a rising process. In test sets on 300s and 420s interval, both $A_s$ and $A_r$ exhibit a growth trend at different degrees, which implies that, under the condition of lower sampling frequency, the temporal and spatial context has less effect on prediction, and route preference ought to become the dominance in map matching. Considering that trajectories of higher sampling frequency are more valuable, it is necessary to ensure matching accuracy of these trajectories as high as possible. According to evaluation results on different sampling interval, the value of $\alpha$ is set to 0.7.

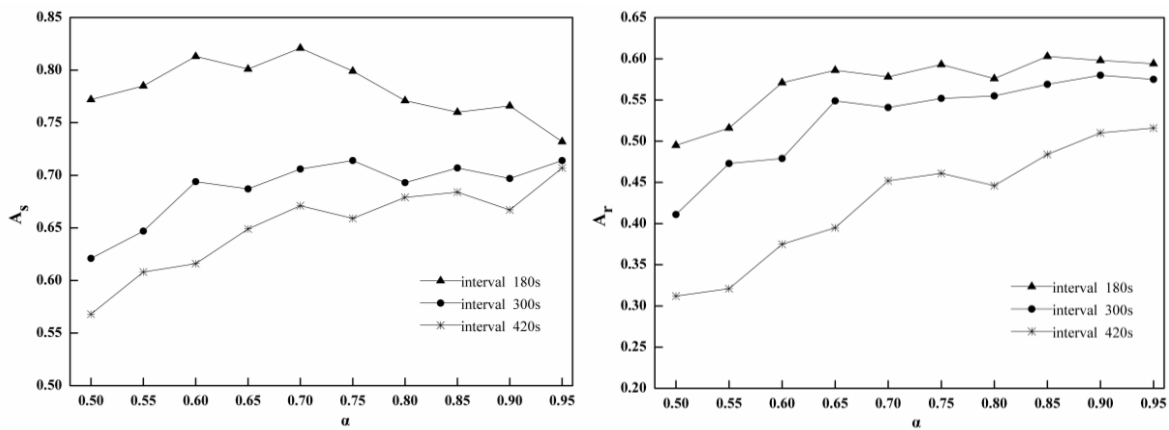

Figure 8 Evaluation of $\alpha$ under different sampling interval

When sampling interval exceeds 180s, $A_s$ and $A_r$ drop lower than 80% and 50% respectively. This implies that $CRF^1$, which depends on temporal-spatial context, can hardly maintain satisfying effectiveness of map matching. So we integrate route preference into CRF-based algorithm to provide replenishment of context. Next, we compare our algorithm combining CRF[1] and RPM (CRF[1]+RPM) with CRF[1]. As presented in figure 9, with sampling interval increasing, $A_s$ and $A_r$ of these two algorithms drop at different degrees. However, comparing with CRF[1], $A_s$ of CRF[1]+RPM drops slowly and tends to be stabilized. This means that the improvement of $A_s$, which is attributed to route preference mining, are more significant with sampling interval increasing, and the growth rate of $A_r$ is more stable and still exceeds 12%. This part of experiment indicates that CRF[1]+RPM outperform CRF[1] for all sampling rate significantly. Further, our algorithm tends to correctly match or mismatch an entire trajectory. Once mismatching of an observation occurs, a considerable proportion of observations would be matched to incorrect road segments. Therefore, the growth of $A_r$ is beyond that of $A_s$. Figure 10 presents a comparison of matching results of CRF[1] and CRF[1]+RPM. As showed in Figure 10, a trajectory with 300s sampling interval, consisting of four GPS observations, is matched to wrong path using CRF[1], while we get the correct path using CRF[1]+RPM.

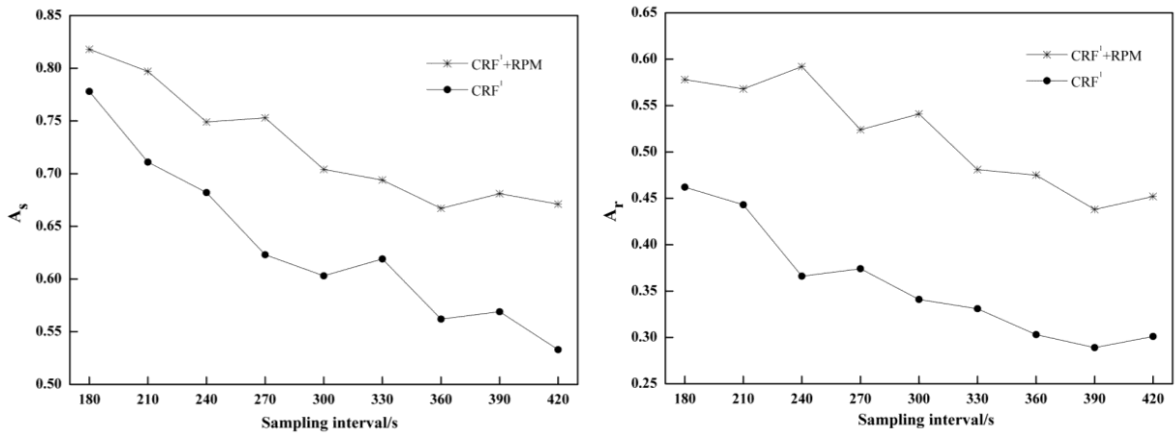

Figure 9 the results of CRF[1] and CRF+RPM

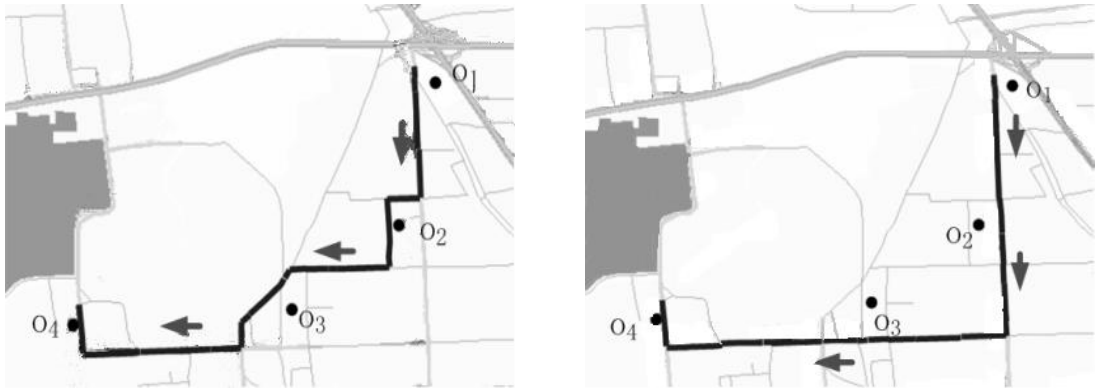

Figure 10 matching results of CRF[1] before (left) and after (right) integrating route preference mining

Finally, we validate the influence of the accumulation of data amount on the effectiveness. Similar to above, our algorithm is evaluated at different sampling intervals: 180s, 300s and 420s, and the amount of trajectory used for route preference analysis is increased successively with incremental of one month. Figure 11 presents the evaluation results. In first two months, the proportion of repeated path is so low that route preference mining can hardly make a great contribution to improve performance, since this stage are mainly in the period of generating rare paths. When data amount is accumulated to 3 months, accuracy begins to grow sharply following the rise of repeated path ratio. And when data amount reaches to 5 months, $A_s$ and $A_r$ are improved by nearly 10%+ and 15%+ respectively. In addition, we can also find that RPM is more effective for the low-sampling-rate trajectories. Note that, the trajectories we used are derived from taxis trips, which are less regular due to randomness of picking up passengers. If this technology is applied to private cars, it can be inferred that higher performance is achieved.

# 6 Conclusions

This paper proposes a CRF-based map-matching algorithm, which can employ the advantages of integrating the context information into features flexibly. The spatial and temporal correlations between neighborhood sampling points are chosen as features in order to improve distinguishability of the trajectory. To further improve the algorithm performance, in the case of low sampling rate, we discover and utilize the personal route preference information to supply the lack of effective features. Experimental results illustrate that this algorithm can accurately distinguish the actual path from other paths on multiple sampling frequency data sets. Compared with the other map matching method, the performance is significantly improved.

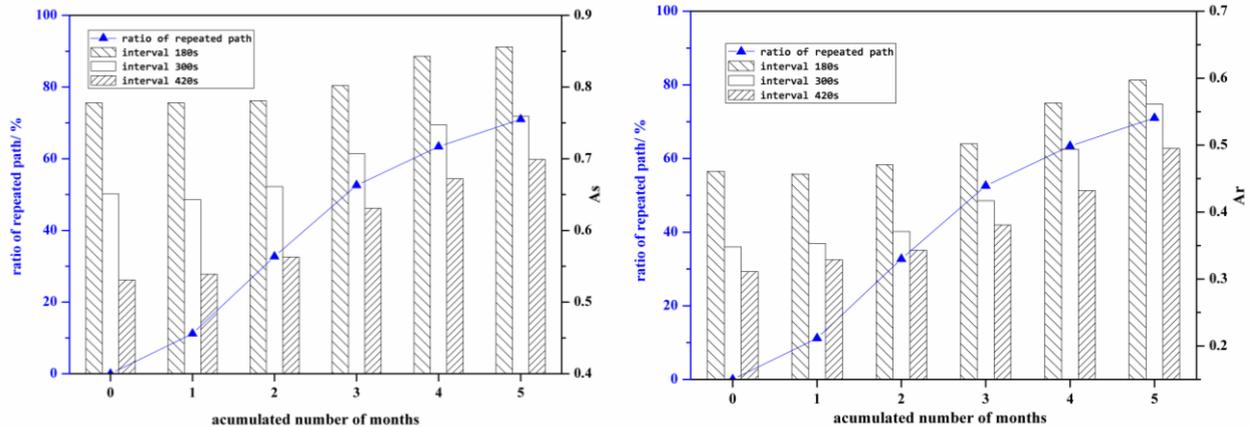

Figure 11 the evaluation of historical accumulation impact on accuracy